\documentclass[a4paper,11pt]{article}
\usepackage{pos}

\newcommand{\gsim}{\;\rlap{\lower 3.5 pt \hbox{$\mathchar \sim$}} \raise 1pt
 \hbox {$>$}\;}
\newcommand{\lsim}{\;\rlap{\lower 3.5 pt \hbox{$\mathchar \sim$}} \raise 1pt
 \hbox {$<$}\;}

\title{Progress in two-loop electroweak corrections to $gg\to HH$ and $gg \to g H$}
\author*[a]{Hantian Zhang}
\affiliation[a]{
Institut f{\"u}r Theoretische Teilchenphysik,
    Karlsruhe Institute of Technology (KIT),\\
  Wolfgang-Gaede Strasse 1, 76128 Karlsruhe, Germany

}

\emailAdd{hantian.zhang@kit.edu}
\abstract{
In these proceedings, we summarise our recent calculations of next-to-leading order electroweak corrections to Higgs boson pair and Higgs boson plus jet production \cite{Davies:2022ram,Davies:2023npk}.
The calculations are divided into different regions.
In the high-energy region, 
we analytically calculate the Higgs boson contribution to the leading two-loop Yukawa corrections for $gg\to HH$.
These corrections are generated by a single virtual Higgs boson exchange within the top quark loop.
Our high-energy expansion yields precise predictions for the region where the Higgs boson transverse momenta $p_T > 120 $ GeV.
In the low-energy region, %$\sqrt{s} < 300$ GeV,
we compute the complete two-loop electroweak corrections to $gg\to HH$ and $gg \to g H$.
We obtain analytic results
through the large top quark mass expansion, covering all sectors of the Standard Model.
}

\FullConference{The European Physical Society Conference on High Energy Physics (EPS-HEP2023)\\
 21-25 August 2023\\
Hamburg, Germany\\}

\begin{document}
\maketitle

\section{Introduction}
The precise study of spontaneously electroweak (EW) symmetry breaking mechanism of the Standard Model (SM) 
is one of the primary targets of the Large Hadron Collider (LHC) programme at CERN.
In this context, Higgs boson pair and Higgs boson plus jet production are two key processes at the LHC, 
which have the further potential to reveal new physics effects beyond the SM (BSM). 
One important feature of the $2\to 2$ Higgs boson production processes
is the access to the Higgs boson transverse momenta ($p_T$) spectrum, which is known to be sensitive to new physics effects 
%\cite{LHCHiggsCrossSectionWorkingGroup:2016ypw,Greljo:2017spw}.
\cite{Greljo:2017spw}.
Moreover, Higgs boson pair production enjoys an additional feature in probing 
the Higgs self-coupling,
which controls the shape of the Higgs potential for the EW symmetry breaking.
The determination of the Higgs self-coupling is one of the most important tasks in the upcoming high-luminosity (HL) phase of the LHC.
Precise predictions for Higgs boson pair production within the SM is
the crucial ingredient for the
examination of the EW symmetry breaking mechanism
and the detection of subtle effects from BSM scenarios \cite{
%Plehn:1996wb,Djouadi:1999rca,
Abouabid:2021yvw,Iguro:2022fel}.

For Higgs boson pair production at the LHC, 
the gluon-fusion $gg\to HH$ process represents the major production channel.
For this process, the next-to-leading order (NLO) QCD corrections with full top quark mass $m_t$ dependence have been calculated in \cite{Borowka:2016ehy,Borowka:2016ypz,Baglio:2018lrj,Bonciani:2018omm,Davies:2019dfy}.
These QCD calculations involving virtual top quarks are difficult. 
The successful approaches  for tackling this problem are numerical approaches~\cite{Borowka:2016ehy,Baglio:2018lrj,Borowka:2016ypz} 
and analytic approximations in complementary regions \cite{Bonciani:2018omm, Davies:2018ood,Davies:2018qvx,Davies:2019dfy,Bellafronte:2022jmo,Davies:2023vmj}.  
However, the calculations of NLO EW corrections are even more involved due to more internal mass scales appearing in the loop integrals,
the Higgs self-coupling corrections have been computed in \cite{Borowka:2018pxx}, leading Yukawa-top corrections have been computed in high energy expansion \cite{Davies:2022ram} and large-$m_t$ limit \cite{Muhlleitner:2022ijf}, and recently we have computed the first full EW corrections in the large-$m_t$ expansion~\cite{Davies:2023npk}.

For Higgs boson plus jet production at the LHC, we consider the dominant gluon-fusion  $gg \to g H$ process.
For this process, the NLO QCD corrections with full $m_t$ dependence are known in \cite{Lindert:2018iug,Jones:2018hbb,Chen:2021azt,Bonciani:2022jmb}.
The NLO electroweak corrections via massless bottom quark loops have been computed in \cite{Bonetti:2020hqh}, the corrections induced by a trilinear Higgs coupling in the large-$m_t$ expansion have been calculated in \cite{Gao:2023bll}, and we have computed the first full EW corrections in the large-$m_t$ expansion~\cite{Davies:2023npk}.

In these proceedings, we summarise our progress towards the full NLO EW calculations for $gg\to HH$ and $gg \to gH$ \cite{Davies:2022ram,Davies:2023npk}.

\section{Form factors  for  $gg\to HH$ and $gg \to gH$}
\begin{figure}[hbt!]
  \centering
  \includegraphics[width=\textwidth]{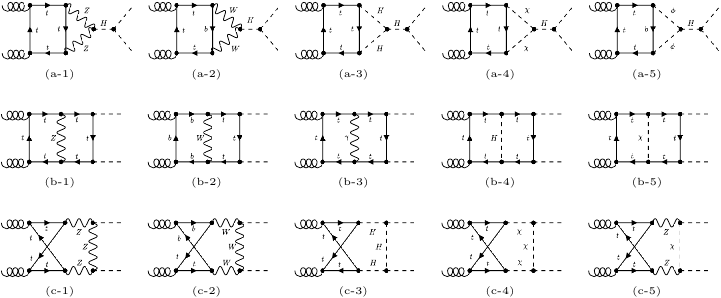}
  \caption{Two-loop Feynman diagrams contributing to $gg\to HH$.
  Dashed, solid, wavy and curly lines correspond to scalar particles,
  fermions, electroweak gauge bosons and gluons, respectively.}
  \label{fig::diags}
\end{figure}

%\subsection{$gg\to HH$}
The amplitude for the process
%\begin{eqnarray}
  $g(q_1)g(q_2)\to H(q_3)H(q_4)$ 
%\end{eqnarray}
can be decomposed into two Lorentz structures
% $A_1^{\mu\nu}$
%and $A_2^{\mu\nu}$ which we define as
\begin{eqnarray}
  A_1^{\mu\nu} = g^{\mu\nu} - {\frac{1}{q_{12}}q_1^\nu q_2^\mu
  }\,, \quad
  A_2^{\mu\nu} = g^{\mu\nu}
                   + \frac{1}{{p_T^2} q_{12}}\left(
                   q_{33}    q_1^\nu q_2^\mu
                   - 2q_{23} q_1^\nu q_3^\mu
                   - 2q_{13} q_3^\nu q_2^\mu
                   + 2q_{12} q_3^\mu q_3^\nu \right)\,, \nonumber
\end{eqnarray}
where $q_{ij} = q_i\cdot q_j$ with $q_1^2=q_2^2=0$ and $q_3^2=q_4^2=m_H^2$,
and $
  p_T =\sqrt{(u\,t-m_H^4)/s} \,
$
is the Higgs boson transverse momentum.
The Mandelstam variables are
$
  s = (q_1+q_2)^2\,, t = (q_1+q_3)^2\,, u = (q_1+q_4)^2\,.
$
We introduce the form factors $F_1$ and $F_2$ as
\begin{eqnarray}
  {\cal M}^{ab} &=& 
  \varepsilon_{1,\mu}\varepsilon_{2,\nu}
  {\cal M}^{\mu\nu,ab}
  \,\,=\,\,
  \varepsilon_{1,\mu}\varepsilon_{2,\nu}
  \delta^{ab} X_0^{\rm ggHH} s 
  \left( F_1 A_1^{\mu\nu} + F_2 A_2^{\mu\nu} \right)
  \,,
       \label{eq::M}
\end{eqnarray}
where $a,b$ are adjoint colour indices,
$X_0^{\rm ggHH} = G_F\alpha_s(\mu) T_F / (2\sqrt{2}\pi)$, $T_F=1/2$, $G_F$
is Fermi's constant and $\alpha_s(\mu)$ is the strong coupling constant
evaluated at the renormalization scale $\mu$.  
In Fig.~\ref{fig::diags} we show sample two-loop diagrams
contributing to $gg\to HH$. 
%
%We decompose the functions
%$F_1$ and $F_2$ introduced in Eq.~(\ref{eq::M}) into ``triangle'' and ``box''
%form factors. $F_1$ has contributions with zero, one and two $s$-channel
%Higgs boson
%propagators whereas $F_2$ only has box contributions. Thus we write
%\begin{eqnarray}
%  F_1 &=& \frac{3 m_H^2}{s-m_H^2}\left(
%          F_{\rm tri}
%          + \frac{m_H^2}{s-m_H^2} \tilde{F}_{\rm tri} 
%          \right)
%          + F_{\rm box1}
%          \,, \nonumber\\
%  F_2 &=& F_{\rm box2}\,.
%                \label{eq::F_12}
%\end{eqnarray}
%In order to obtain this decomposition it is important to re-write the factors
%of $s$ which occur in the numerators during the calculation using
%$s/(s-m_H^2) = 1 + m_H^2/(s-m_H^2)$.  Note that at two loops $F_{\rm tri}$ is
%not the same as the form factor for single Higgs boson production (as is
%the case for QCD corrections), since loop corrections to the $HHH$ vertex also
%enter here.

%At two-loop order we have:
%\begin{itemize}\setlength\itemsep{-.1em}
%\item one-particle irreducible box and triangle diagrams,
%\item one-particle reducible diagrams with a one-loop correction to the $HHH$
%	vertex or a one-loop self-energy correction to the Higgs propagator of a one-loop $gg \to H \to HH$ diagram,
%\item one-loop tadpole corrections to one-loop diagrams.
%\end{itemize}

%At two-loop order there are also contributions without top quarks which are
%not suppressed by small Yukawa couplings.  In these contributions the
%gluons couple to light quarks and the connection to the final-state Higgs
%bosons is mediated via $Z$ bosons.  An example is given by diagram 
%(g-1) in Fig.~\ref{fig::diags} if a light quark runs in
%the fermion loop.  In our expansion these contributions formally contribute to
%the $m_t^0$ term, however in this work we do not compute such diagrams;
%they can be computed following the approach of Ref.~\cite{Bonetti:2020hqh}.

\begin{figure}[hbt!]
  \centering
  \includegraphics[width=\textwidth]{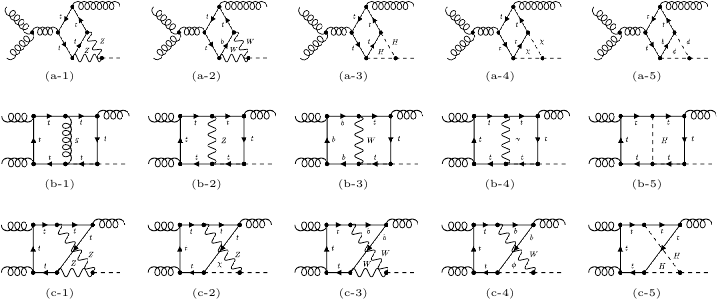}
  \caption{Two-loop Feynman diagrams contributing to $gg\to gH$.
    Dashed, solid, wavy and curly lines correspond to scalar particles,
    fermions, electroweak gauge bosons and gluons, respectively.  
%    Diagrams are also shown which contribute to the NLO QCD corrections.
    }
  \label{fig::diags_gggh}
\end{figure}

%\subsection{$gg\to gH$}
The amplitude for the process 
%\begin{eqnarray}
 $ g(q_1)g(q_2)\to g(q_3)H(q_4)$ 
%\end{eqnarray}
can be decomposed into four physical Lorentz structures
%\footnote{We note that $A_4^{\mu\nu\rho}$ differs from Ref.~\cite{Melnikov:2016qoc} by the factor
%of $1/s$, which we introduce such that all four form factors are dimensionless.}
\begin{eqnarray}
  A_1^{\mu\nu\rho} = g^{\mu\nu} q_2^\rho \,, \qquad  A_2^{\mu\nu\rho} = g^{\mu\rho} q_1^\nu \,, \qquad
  A_3^{\mu\nu\rho} = g^{\nu\rho} q_3^\mu \,, \qquad  
                     A_4^{\mu\nu\rho} = \frac{1}{s} q_3^\mu q_1^\nu q_2^\rho \,. \nonumber
\end{eqnarray}
The corresponding four form factors are defined through
\begin{eqnarray}
  {\cal M}^{abc} &=& f^{abc} X_0^{\rm gggH} 
                     \varepsilon_{1,\mu}\varepsilon_{2,\nu} \varepsilon_{3,\rho}
                     \sum_{i=1}^{4} F_i A_i^{\mu\nu\rho}
  \,,
  \label{eq::M2}
\end{eqnarray}
where 
$
   X_0^{\rm gggH} = 2^{1/4} \sqrt{4\pi\alpha_s(\mu) G_F} \, \frac{\alpha_s(\mu)}{4\pi}\,
$,
and $c$ is the adjoint colour index of the final-state gluon.
The Mandelstam variables are defined as in  $gg \to HH$,
apart from here we have $q_3^2=0$ and $p_T = \sqrt{u\,t/s}$.  Sample Feynman diagrams for
$gg\to gH$ are given in Fig.~\ref{fig::diags_gggh}.  

%The classification is
%similar to $gg\to HH$, we again include all one-particle reducible and all
%tadpole contributions.
%%
%Note that for the QCD corrections, we also include the one-loop self-energy corrections to the gluon propagators
%and the one-loop vertex corrections to the triple-gluon vertex of the one-loop diagrams. The corrections to the quartic-gluon vertex do not appear at the two-loop order of this process.

We define the perturbative expansion of the form factors as
\begin{eqnarray}
  F &=& F^{(0)} + \frac{\alpha_s(\mu)}{\pi} F^{(1,0)} + \frac{\alpha}{\pi} F^{(0,1)} 
        + \cdots
  \,,
  \label{eq::F}
\end{eqnarray}
where $\alpha$ is the fine structure constant and 
the ellipses indicate higher-order QCD and EW corrections.

%%%%%%%%%%%%%%%%%%%%%%%%%%%%%%%%%%%%%%
\section{Leading Yukawa corrections to $gg\to HH$ in high energy expansion}
As a concrete example, we present the details for the calculation of the leading Yukawa corrections to the Higgs-exchange two-loop box diagram (b-4) shown in Fig.~\ref{fig::diags}.
%
%For the computation of the two-loop integrals we follow two approaches, which
%we describe in the following.  For this purpose it is convenient to distinguish
%the mass of the final-state Higgs bosons ($m_H^{\rm ext}$) from that of the Higgs
%boson which propagates in the loops ($m_H^{\rm int}$).
%This means that for the process
%$gg\to HH$ we have the following dimensionful quantities: the Mandelstam
%variables $s$ and $t$, and the masses $m_t$, $m_H^{\rm int}$ and
%$m_H^{\rm ext}$.  
For this diagram,
we consider two expansion approaches with the following hierarchies:
\begin{eqnarray}
({\rm A}) \,\; s,t \gg m_t^2 \gg (m_H^{\rm int})^2, (m_H^{\rm ext})^2 \quad \mbox{and} \quad 
({\rm B}) \,\; s,t \gg m_t^2 \approx (m_H^{\rm int})^2 \gg (m_H^{\rm ext})^2\,,
\end{eqnarray}
where $m_H^{\rm int}$ and $m_H^{\rm ext}$ are internal and external Higgs masses respectively.
In approach~(A) we first treat the hierarchy $m_t^2 \gg (m_H^{\rm int})^2$ at the
level of the integrand by applying the hard-mass expansion procedure using {\tt exp}~\cite{Harlander:1998cmq,Seidensticker:1999bb},
and further perform a Taylor expansion in the $m_H^{\rm ext} \to 0$ limit.
For approach (B) we  perform simple Taylor expansions 
for the hierarchy $m_t^2 \approx (m_H^{\rm int})^2 \gg (m_H^{\rm ext})^2$.
At this stage, the problem is reduced to simpler massive two-loop four-point integrals with massless external lines 
that only depend on the variables $s$, $t$ and $m_t$.
We then perform integration-by-parts reductions and derive the system of differential equations for master integrals.
To treat the final  hierarchy $s,t \gg m_t^2$, we construct the asymptotic expansion at the level of the master integrals by inserting a power-log ansatz
\begin{eqnarray}
  I_{n} &=& \sum\limits_{i=-3}^{i_{max}} \sum\limits_{j=-4}^{j^+} \sum\limits_{k=0}^{i+4} c_{ijk}^{(n)}(s,t) \, \epsilon^i \, {m}_t^{2j} \, \log^k({m}_t^2) \,,
            \label{eq::ansatz}
\end{eqnarray}
into the system of differential equations w.r.t. $m_t$.
We then solve the expanded differential equations
% Since there are no spurious poles in $\epsilon$
%either in the physical amplitudes or the differential equations we can choose
%$i_{max}=0$ for all master integrals. 
to a high order in $m_t$ in terms of unknown boundary conditions.
The boundary conditions need to be calculated analytically 
from the corresponding master integrals in
the limit ${m}_t \to 0$.  
We
employ the asymptotic expansion method~\cite{Beneke:1997zp} using \texttt{asy.m}~\cite{Pak:2010pt} to
obtain integral representations for the required boundary conditions.  
These integrals
are subsequently solved using Mellin-Barnes techniques
together with our in-house package \texttt{AsyInt}.
Finally we obtain analytic results for the amplitudes expanded up to order $m_t^{120}$.
For the numerical evaluation, we employ the Pad\'{e} improved approximation to enlarge the radius of convergence of our results.
%We conclude that the inclusion of the quartic terms in $m_H$ and
%cubic terms in $\delta$ provides an approximation to the (unknown)
%exact result below the percent level (see also Fig.~2 or Ref.~\cite{Chen:2022rua}
%which shows a comparison for $gg\to ZH$).
\begin{figure}[hbt!]
  \centering
  \begin{tabular}{cc}
    \includegraphics[width=0.45\textwidth]{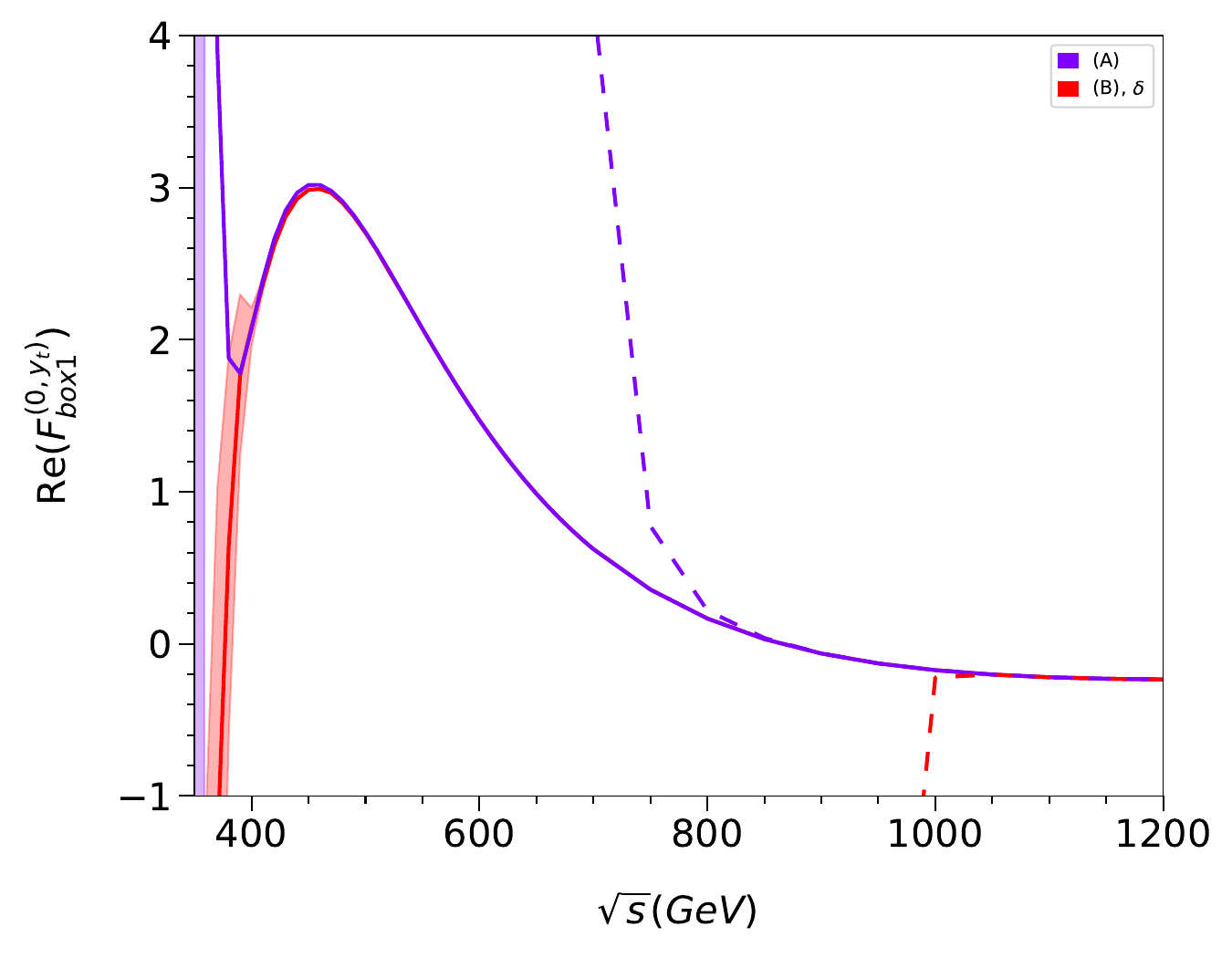}
    &
    \includegraphics[width=0.45\textwidth]{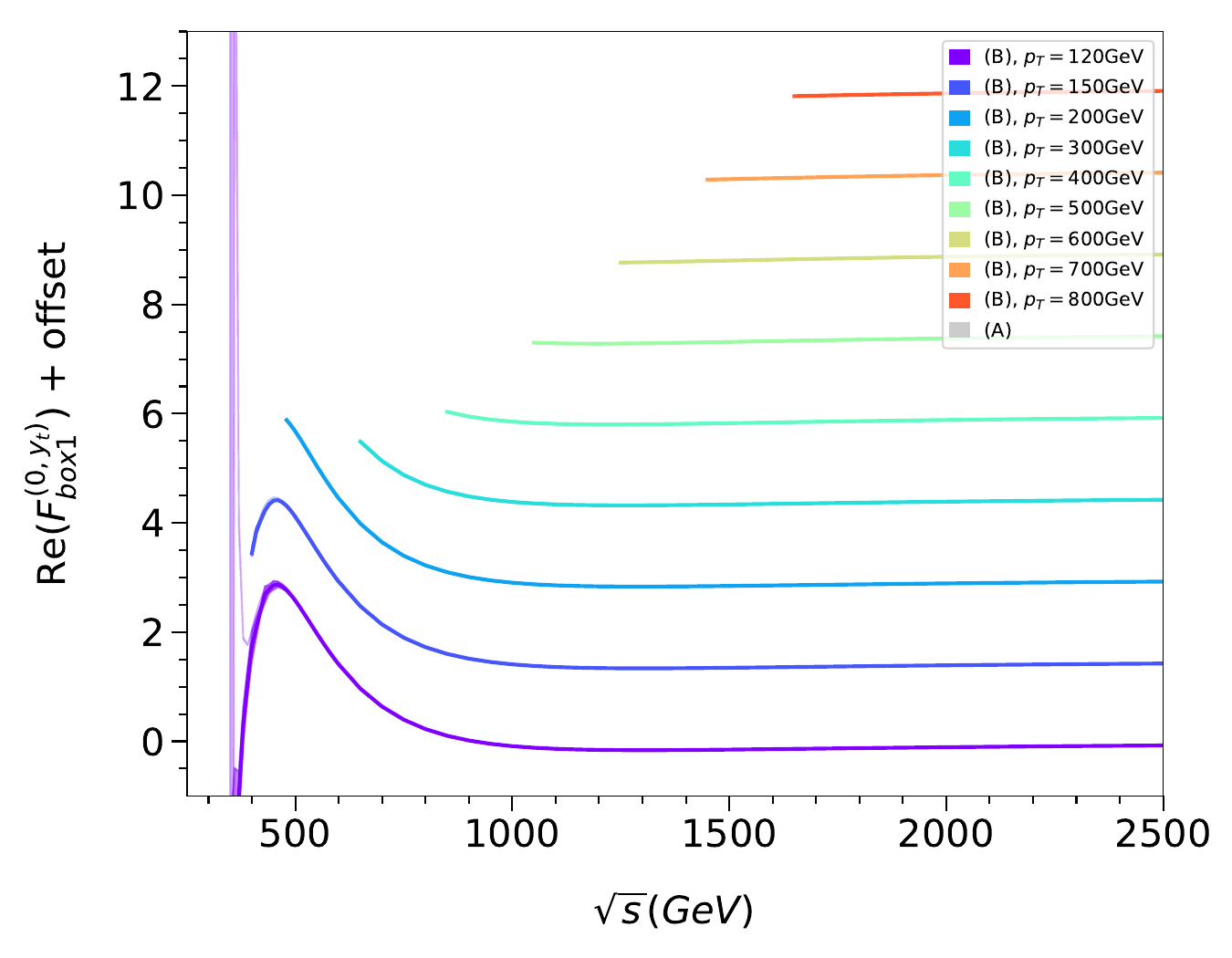}
  \end{tabular}
  \caption{\label{fig::F12}
  Real part of $F_{\rm box1}$ for fixed scattering angle $\theta=\pi/2$ (left) and different fixed $p_T$ values (right). 
  For the fixed $p_T$ plot, an offset is applied 
    such that the curves for the different $p_T$ values are separated.
    No offset is used for the lowest $p_T$ value.
%    We shift the subsequent $p_T$ curves by 1.5 to the positive $y$ axis.
%    The coloured curves and the corresponding bands correspond to 
%    approach~(B). Results for approach~(A) are only shown as faint uncertainty
%    bands.
    For $p_T\ge120$~GeV the central values of approach~(A) and~(B) agree.
    }
\end{figure}

Here we show results for the real part of box-type form factor of $F_{1}$ for
fixed transverse momentum $p_T$ and fixed scattering angle in Fig.~\ref{fig::F12}.
For the fixed scattering angle plot,
the solid curves represent Pad\'e results with uncertainty bands\footnote{We employ the so-called pole-distance re-weighted Pad\'e approximants and the corresponding uncertainties~\cite{Davies:2020lpf}.} and the dashed curves show naive expansions.
We observe that the naive expansions start to diverge for $\sqrt{s}\approx 800$~GeV for
approach~(A) and for $\sqrt{s}\approx 1000$~GeV for approach~(B), 
while the central value of Pad\'e results agree between both approaches down to $\sqrt{s}\approx 400$~GeV.
% we show the real and
%imaginary parts of $F_{\rm box1}$ and $F_{\rm box2}$ for $p_T$ between
%$120$~GeV and $800$~GeV. 
%In all cases we show the Pad\'e-improved
%results as solid curves and the naive expansion as dashed lines. 
%
For the fixed $p_T$ plot, 
the coloured curves correspond to the results from approach~(B), and the results from approach~(A) 
are only shown as faint uncertainty bands. %They are only visible for small
%values of $p_T$ where one observes deviations between the two approaches.
These curves show that both expansion approaches
yield equivalent physical results.

%%%%%%%%%%%%%%%
\section{Full EW corrections to $gg\to HH$ and $gg\to gH$ in large-$m_t$ expansion}
We perform calculations for the full EW corrections to $gg\to HH$ and $gg\to gH$ 
through the large-$m_t$ expansion.
To this end, we assume the following hierarchy 
\begin{eqnarray}
  m_t^2 \gg \xi_W m_W^2, \xi_Z m_Z^2 \gg s,t,m_W^2,m_Z^2, m_H^2 \,,
  \label{eq::hierarchy-xi}
\end{eqnarray}
where $\xi_Z$, $\xi_W$ are the general gauge parameters for the $Z$ and $W$ bosons,
and perform the large-$m_t$ expansion with {\tt exp}.
Through this procedure, we obtain analytic results for the bare two-loop amplitudes up to order $1/m_t^{4}$ in the $R_\xi$ gauge
and order $1/m_t^{10}$ $(1/m_t^8)$ in the Feynman gauge for $gg\to HH$ $(gg\to gH)$.
For the renormalisation, 
we express our one-loop amplitudes
in terms of independent parameters
$
  \{ e, m_W, m_Z, m_t, m_H \}
$
with $e = \sqrt{4\pi\alpha}$, and introduce one-loop on-shell counterterms.
We also renormalise the wave function of the external Higgs boson in the on-shell scheme.
Note that
tadpole contributions are included in all parts of our calculations. 
%This guarantees that the top quark mass
%counterterm is gauge-parameter independent.  
%This prescription is equivalent to the so-called {\it Fleischer–Jegerlehner tadpole scheme}.
After renormalisation, $\xi_W$ and $\xi_Z$ drop out for both $gg\to HH$ and $gg\to gH$ amplitudes.

\begin{figure}[hbt!]
  \centering
  \begin{tabular}{cc}
    \includegraphics[width=.40\textwidth]{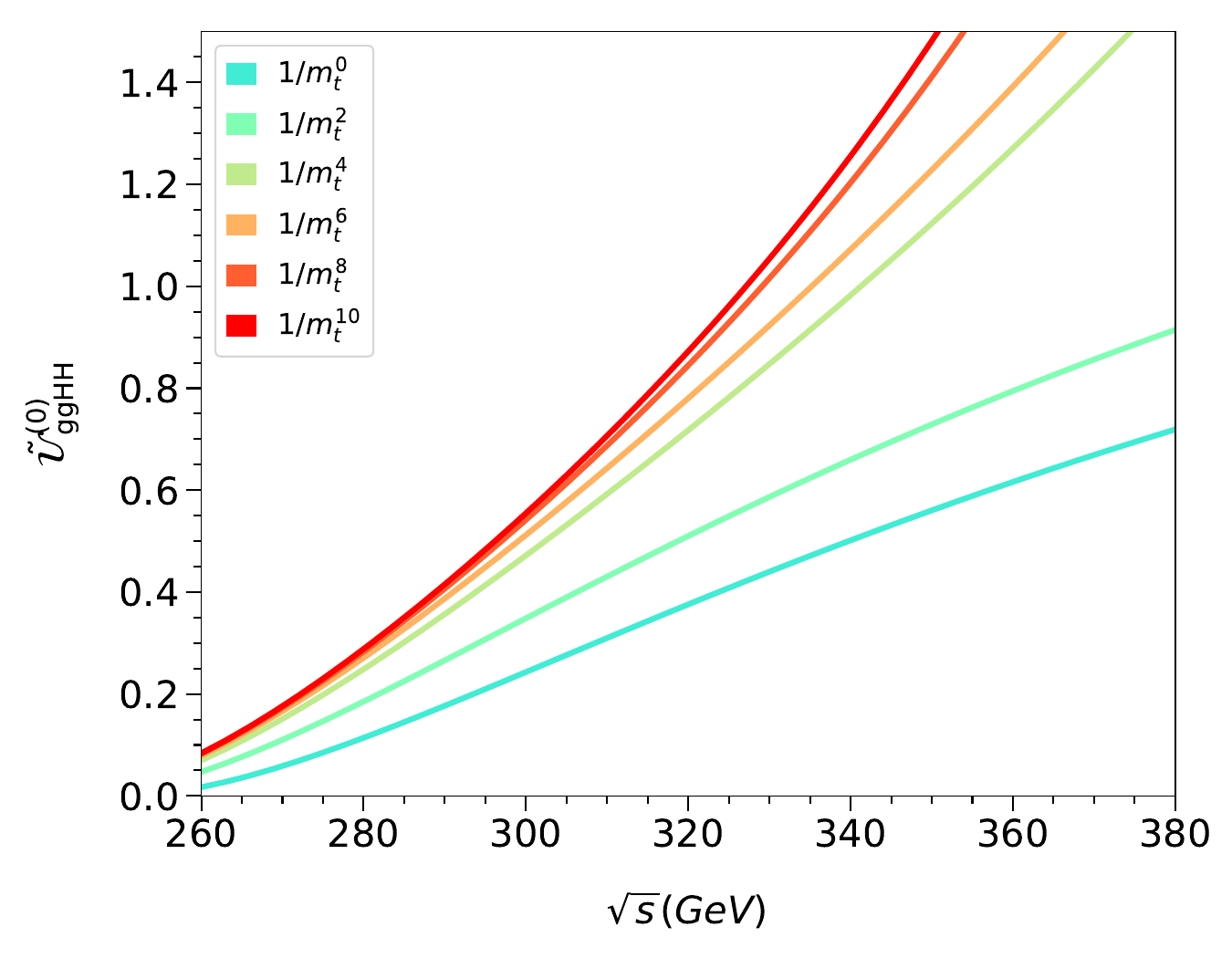} &
    \includegraphics[width=.40\textwidth]{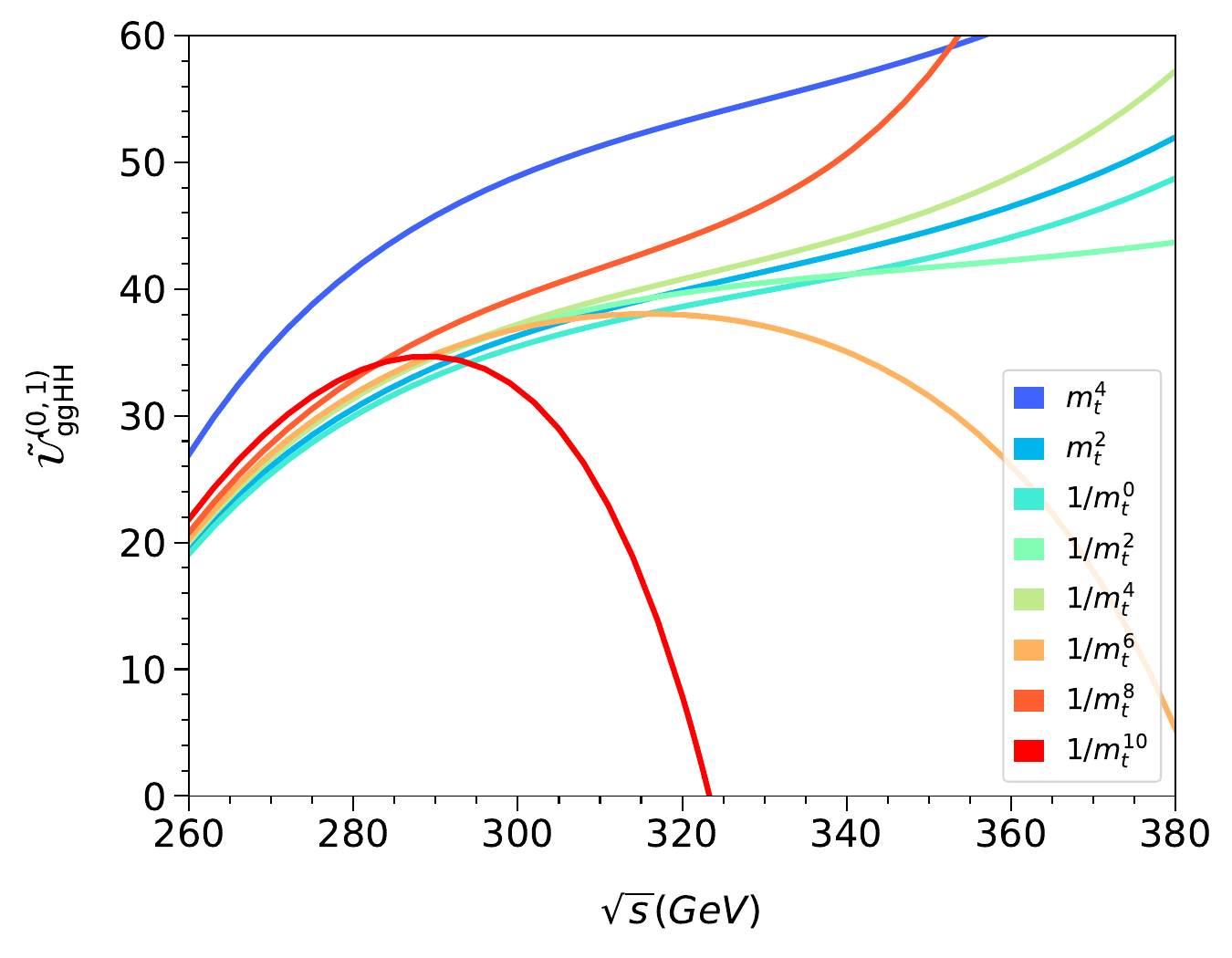}
  \end{tabular}
  \caption{\label{fig::gghh}
      LO $\tilde{{\cal U}}_{\rm ggHH}^{(0)}$ (left) and NLO EW $\tilde{{\cal U}}_{\rm ggHH}^{(0,1)}$ (right) matrix elements plotted as a function of
      $\sqrt{s}$ for $gg\to HH$. Results are shown up to order $1/m_t^{10}$.
       }
\end{figure}

\begin{figure}[htb!]
  \centering
  \begin{tabular}{cc}
    \includegraphics[width=.40\textwidth]{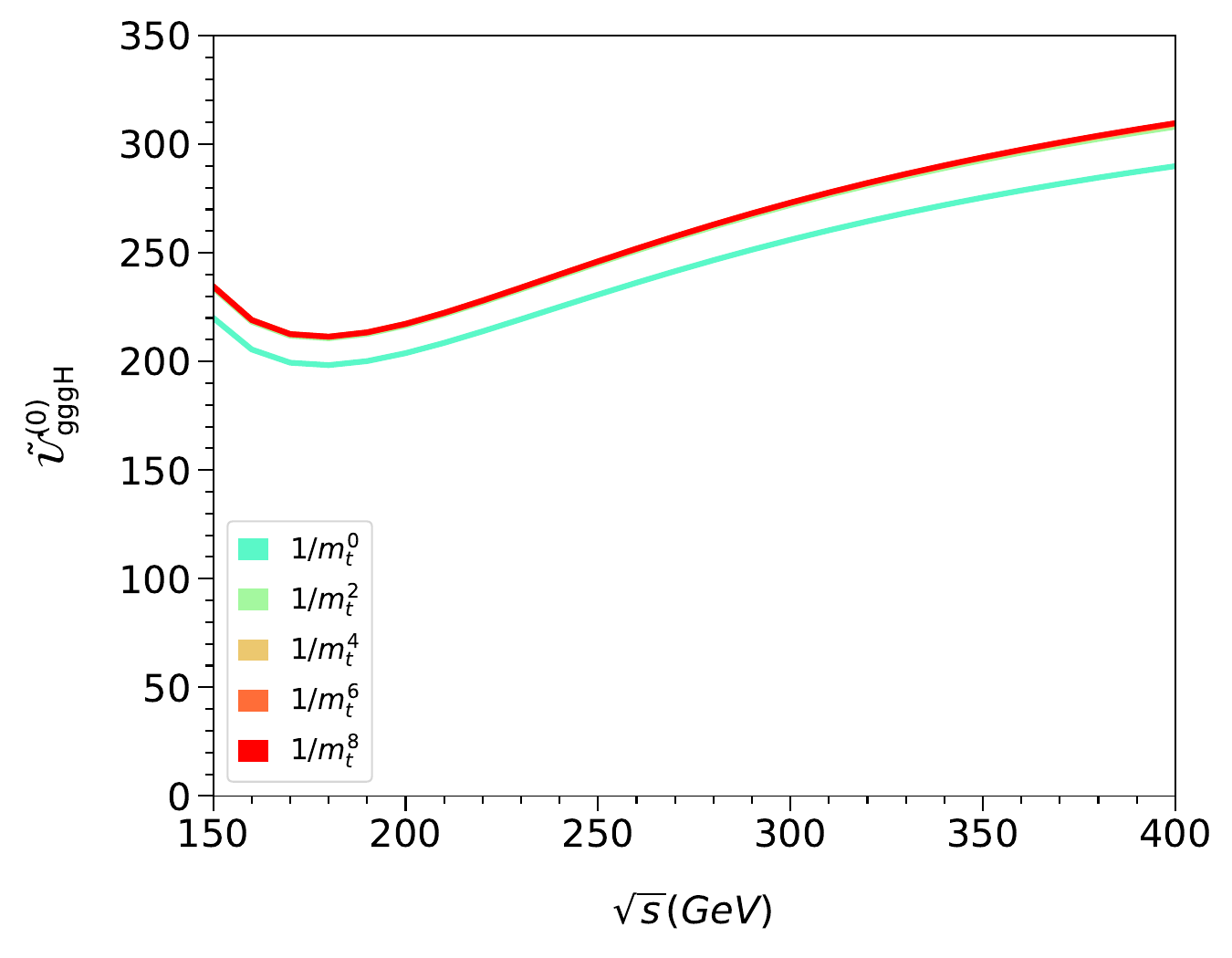} &
    \includegraphics[width=.40\textwidth]{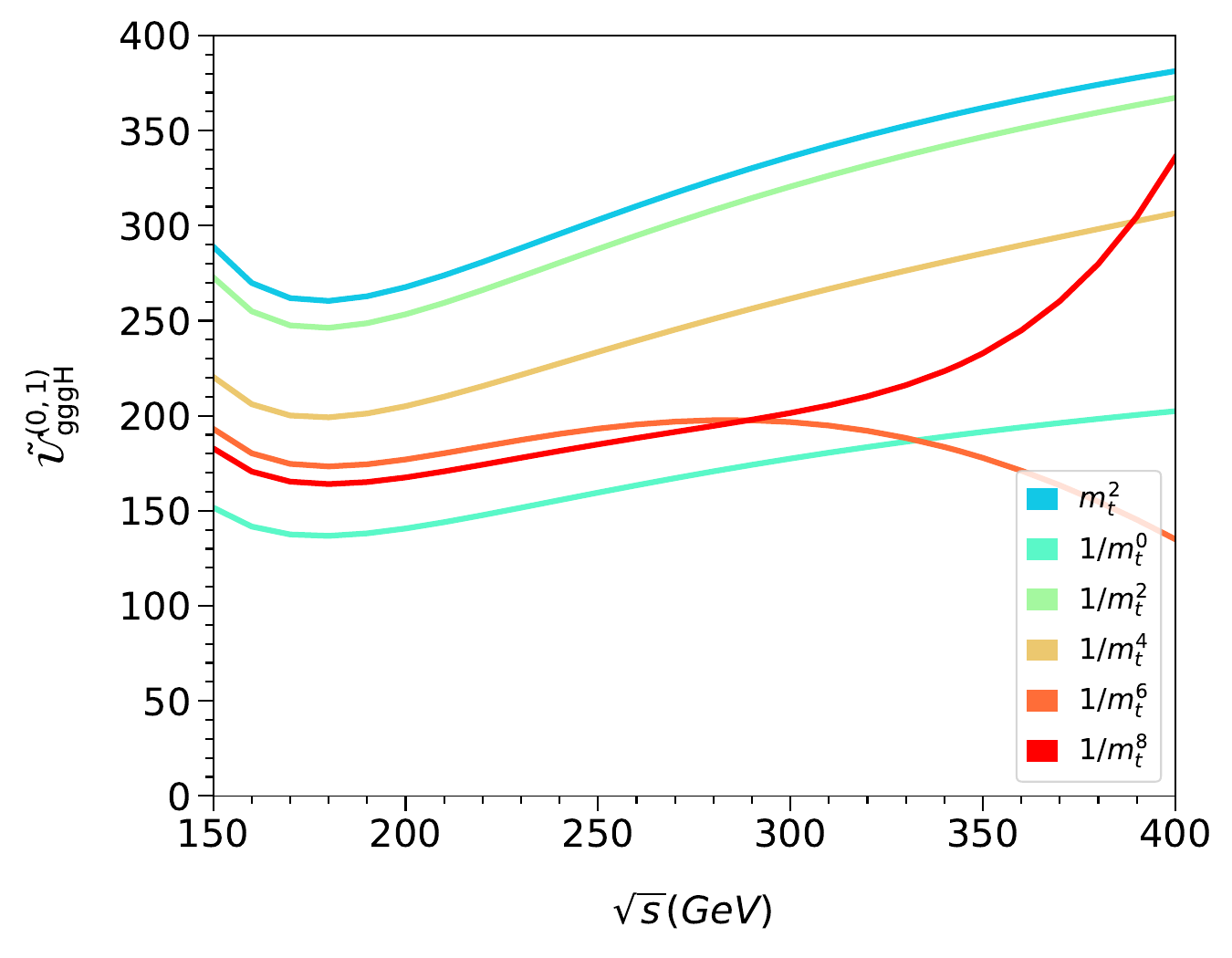}
  \end{tabular}
  \caption{\label{fig::gggH}
  LO $\tilde{{\cal U}}_{\rm gggH}^{(0)}$ (left) and NLO EW $\tilde{{\cal U}}_{\rm gggH}^{(0,1)}$ (right) matrix elements plotted as a function of
      $\sqrt{s}$ for $gg\to gH$. Results are shown up to order $1/m_t^{8}$.
    }
\end{figure}

%%%%%%%%%%%%%%%%
%\subsection{Results for $gg\to HH$}
For the numerical evaluation, we adopt the $G_\mu$
scheme and use the input values
$
  m_t = 172~\mbox{GeV}\,,  m_H = 125~\mbox{GeV}\,,
  m_W = 80~\mbox{GeV}\,, m_Z = 91~\mbox{GeV}\,,
$
%Furthermore, we express the form factors in terms of $s$ and $p_T$
and introduce the ratio parameter
$
  \rho_{p_T} = \frac{p_T}{\sqrt{s}}
$.
In the following we choose $\rho_{p_T}=0.1$ and
consider the squared matrix element 
%\begin{eqnarray}
$  {\cal U}_{\rm ggHH} =
       \frac{1}{8^2} \sum_{\rm col} 
       \frac{1}{2^2} \sum_{\rm pol} |{\cal M}^{ab}|^2 
%  = \frac{1}{16}
%  \left( X_0^{\rm ggHH} s \right)^2 \left( |F_1|^2 + |F_2|^2 \right)
  = \frac{1}{16} 
  \big( X_0^{\rm ggHH} s \big)^2 \tilde{ {\cal U} }_{\rm ggHH}  $
 for $gg\to HH$ and
$  {\cal U}_{\rm gggH} =
   \frac{1}{8^2} \sum_{\rm col} 
   \frac{1}{2^2} \sum_{\rm pol} |{\cal M}^{abc}|^2
  =  \frac{3}{32} 
      { \big( X_0^{\rm gggH} \big)^2 s \,\, 
      \tilde{ {\cal U} }_{\rm gggH} }$
 for $gg\to gH$.
%\end{eqnarray}
%For the numerical evaluation of the massive two- and three-point functions
%we use the program \texttt{Package-X}~\cite{Patel:2016fam}.
The LO and NLO EW numerical results are shown in Fig.~\ref{fig::gghh} for $gg\to HH$
and Fig.~\ref{fig::gggH} for $gg\to gH$ with different expansion order in $m_t$.
For both processes, our large-$m_t$ expansions yield reasonable predictions for the $\sqrt{s} \lsim 290$~GeV region.
We observe that the NLO EW corrections for $gg\to HH$ can be sizeable, potentially reaching $\mathcal{O}(10\%)$ compared to the LO,
while they are small for $gg\to gH$.
The analytic NLO QCD results for $gg\to gH$ are also available in \cite{Davies:2023npk}.

%\vspace{-2cm}

\section{Conclusion}
In these proceedings, we have summarised our recent analytic calculations of two-loop EW corrections to $gg\to HH$ and $gg\to gH$.
We have presented the first NLO leading Yukawa corrections to $gg \to HH$ in the high energy limit.
%In this context, two different expansion approaches are proposed to tackle the challenging two-loop four-point amplitudes with multiple internal masses.
We have proposed two different expansion approaches to tackle this problem and shown that they yield physically equivalent and precise predictions even for Higgs boson $p_T$ as small as 120~GeV.
We have also presented the first full NLO EW corrections to both $gg\to HH$ and $gg\to gH$ in the large-$m_t$ expansion,
including all sectors of the SM.
%Our analytic results are obtained up to expansion order $1/m_t^{10}$ for $gg\to HH$ and $1/m_t^{8}$ for $gg\to gH$.
Our results also shown that the NLO EW corrections to the $gg\to HH$ can  potentially be sizeable.

\section*{Acknowledgments}
I would like to thank Joshua Davies, Go Mishima, Kay Sch\"onwald and Matthias Steinhauser for close collaboration on the projects in this contribution. 
This research was supported by the Deutsche Forschungsgemeinschaft (DFG, German Research Foundation) under grant 396021762 --- TRR 257 “Particle Physics Phenomenology after the Higgs Discovery”.

\footnotesize
\providecommand{\href}[2]{#2}\begingroup\raggedright\endgroup

%
%\footnotesize
%\bibliographystyle{JHEP}
%%\bibliographystyle{apsrev4-2}
%\bibliography{reference}

\begin{thebibliography}{10}

\bibitem{Davies:2022ram}
J.~Davies, G.~Mishima, K.~Sch\"onwald, M.~Steinhauser and H.~Zhang,
%  \emph{{Higgs boson contribution to the leading two-loop Yukawa corrections to
%  gg \textrightarrow{} HH}},
  \href{https://doi.org/10.1007/JHEP08(2022)259}{\emph{JHEP} {\bfseries 08}
  (2022) 259} [\href{https://arxiv.org/abs/2207.02587}{{\ttfamily
  2207.02587}}].

\bibitem{Davies:2023npk}
J.~Davies, K.~Sch\"onwald, M.~Steinhauser and H.~Zhang, 
%\emph{{Next-to-leading
%  order electroweak corrections to $gg \to HH$ and $gg \to gH$ in the
%  large-$m_t$ limit}},
  \href{https://doi.org/10.1007/JHEP10(2023)033}{\emph{JHEP} {\bfseries 10}
  (2023) 033} [\href{https://arxiv.org/abs/2308.01355}{{\ttfamily
  2308.01355}}].

\bibitem{Greljo:2017spw}
A.~Greljo, G.~Isidori, J.~M. Lindert, D.~Marzocca and H.~Zhang,
%  \emph{{Electroweak Higgs production with HiggsPO at NLO QCD}},
  \href{https://doi.org/10.1140/epjc/s10052-017-5422-4}{\emph{Eur. Phys. J. C}
  {\bfseries 77} (2017) 838}
  [\href{https://arxiv.org/abs/1710.04143}{{\ttfamily 1710.04143}}].

\bibitem{Abouabid:2021yvw}
H.~Abouabid, A.~Arhrib, D.~Azevedo, J.~E. Falaki, P.~M. Ferreira,
  M.~M\"uhlleitner et~al., 
%  \emph{{Benchmarking di-Higgs production in various
%  extended Higgs sector models}},
  \href{https://doi.org/10.1007/JHEP09(2022)011}{\emph{JHEP} {\bfseries 09}
  (2022) 011} [\href{https://arxiv.org/abs/2112.12515}{{\ttfamily
  2112.12515}}].

\bibitem{Iguro:2022fel}
S.~Iguro, T.~Kitahara, Y.~Omura and H.~Zhang, 
%\emph{{Chasing the two-Higgs
%  doublet model in the di-Higgs boson production}},
  \href{https://doi.org/10.1103/PhysRevD.107.075017}{\emph{Phys. Rev. D}
  {\bfseries 107} (2023) 075017}
  [\href{https://arxiv.org/abs/2211.00011}{{\ttfamily 2211.00011}}].

\bibitem{Borowka:2016ehy}
S.~Borowka, N.~Greiner, G.~Heinrich, S.~P. Jones, M.~Kerner, J.~Schlenk et~al.,
%  \emph{{Higgs Boson Pair Production in Gluon Fusion at Next-to-Leading Order
%  with Full Top-Quark Mass Dependence}},
  \href{https://doi.org/10.1103/PhysRevLett.117.079901}{\emph{Phys. Rev. Lett.}
  {\bfseries 117} (2016) 012001}
  [\href{https://arxiv.org/abs/1604.06447}{{\ttfamily 1604.06447}}].

\bibitem{Borowka:2016ypz}
S.~Borowka, N.~Greiner, G.~Heinrich, S.~P. Jones, M.~Kerner, J.~Schlenk et~al.,
%  \emph{{Full top quark mass dependence in Higgs boson pair production at
%  NLO}}, 
  \href{https://doi.org/10.1007/JHEP10(2016)107}{\emph{JHEP} {\bfseries
  10} (2016) 107} [\href{https://arxiv.org/abs/1608.04798}{{\ttfamily
  1608.04798}}].

\bibitem{Baglio:2018lrj}
J.~Baglio, F.~Campanario, S.~Glaus, M.~M\"uhlleitner, M.~Spira and
  J.~Streicher, 
%  \emph{{Gluon fusion into Higgs pairs at NLO QCD and the top
%  mass scheme}},
  \href{https://doi.org/10.1140/epjc/s10052-019-6973-3}{\emph{Eur. Phys. J. C}
  {\bfseries 79} (2019) 459}
  [\href{https://arxiv.org/abs/1811.05692}{{\ttfamily 1811.05692}}].

\bibitem{Bonciani:2018omm}
R.~Bonciani, G.~Degrassi, P.~P. Giardino and R.~Gr\"ober, 
%\emph{{Analytical
%  Method for Next-to-Leading-Order QCD Corrections to Double-Higgs
%  Production}},
  \href{https://doi.org/10.1103/PhysRevLett.121.162003}{\emph{Phys. Rev. Lett.}
  {\bfseries 121} (2018) 162003}
  [\href{https://arxiv.org/abs/1806.11564}{{\ttfamily 1806.11564}}].

\bibitem{Davies:2019dfy}
J.~Davies, G.~Heinrich, S.~P. Jones, M.~Kerner, G.~Mishima, M.~Steinhauser
  et~al., 
%  \emph{{Double Higgs boson production at NLO: combining the exact
%  numerical result and high-energy expansion}},
  \href{https://doi.org/10.1007/JHEP11(2019)024}{\emph{JHEP} {\bfseries 11}
  (2019) 024} [\href{https://arxiv.org/abs/1907.06408}{{\ttfamily
  1907.06408}}].

\bibitem{Davies:2018ood}
J.~Davies, G.~Mishima, M.~Steinhauser and D.~Wellmann, 
%\emph{{Double-Higgs
%  boson production in the high-energy limit: planar master integrals}},
  \href{https://doi.org/10.1007/JHEP03(2018)048}{\emph{JHEP} {\bfseries 03}
  (2018) 048} [\href{https://arxiv.org/abs/1801.09696}{{\ttfamily
  1801.09696}}].

\bibitem{Davies:2018qvx}
J.~Davies, G.~Mishima, M.~Steinhauser and D.~Wellmann, 
%\emph{{Double Higgs
%  boson production at NLO in the high-energy limit: complete analytic
%  results}}, 
  \href{https://doi.org/10.1007/JHEP01(2019)176}{\emph{JHEP}
  {\bfseries 01} (2019) 176}
  [\href{https://arxiv.org/abs/1811.05489}{{\ttfamily 1811.05489}}].

\bibitem{Bellafronte:2022jmo}
L.~Bellafronte, G.~Degrassi, P.~P. Giardino, R.~Gr\"ober and M.~Vitti,
%  \emph{{Gluon fusion production at NLO: merging the transverse momentum and
%  the high-energy expansions}},
  \href{https://doi.org/10.1007/JHEP07(2022)069}{\emph{JHEP} {\bfseries 07}
  (2022) 069} [\href{https://arxiv.org/abs/2202.12157}{{\ttfamily
  2202.12157}}].

\bibitem{Davies:2023vmj}
J.~Davies, G.~Mishima, K.~Sch\"onwald and M.~Steinhauser, 
%\emph{{Analytic
%  approximations of 2 \textrightarrow{} 2 processes with massive internal
%  particles}}, 
  \href{https://doi.org/10.1007/JHEP06(2023)063}{\emph{JHEP}
  {\bfseries 06} (2023) 063}
  [\href{https://arxiv.org/abs/2302.01356}{{\ttfamily 2302.01356}}].

\bibitem{Borowka:2018pxx}
S.~Borowka, C.~Duhr, F.~Maltoni, D.~Pagani, A.~Shivaji and X.~Zhao,
%  \emph{{Probing the scalar potential via double Higgs boson production at
%  hadron colliders}},
  \href{https://doi.org/10.1007/JHEP04(2019)016}{\emph{JHEP} {\bfseries 04}
  (2019) 016} [\href{https://arxiv.org/abs/1811.12366}{{\ttfamily
  1811.12366}}].

\bibitem{Muhlleitner:2022ijf}
M.~M\"uhlleitner, J.~Schlenk and M.~Spira, 
%\emph{{Top-Yukawa-induced
%  corrections to Higgs pair production}},
  \href{https://doi.org/10.1007/JHEP10(2022)185}{\emph{JHEP} {\bfseries 10}
  (2022) 185} [\href{https://arxiv.org/abs/2207.02524}{{\ttfamily
  2207.02524}}].

\bibitem{Lindert:2018iug}
J.~M. Lindert, K.~Kudashkin, K.~Melnikov and C.~Wever, 
%\emph{{Higgs bosons with
%  large transverse momentum at the LHC}},
  \href{https://doi.org/10.1016/j.physletb.2018.05.009}{\emph{Phys. Lett. B}
  {\bfseries 782} (2018) 210}
  [\href{https://arxiv.org/abs/1801.08226}{{\ttfamily 1801.08226}}].

\bibitem{Jones:2018hbb}
S.~P. Jones, M.~Kerner and G.~Luisoni, 
%\emph{{Next-to-Leading-Order QCD
%  Corrections to Higgs Boson Plus Jet Production with Full Top-Quark Mass
%  Dependence}},
  \href{https://doi.org/10.1103/PhysRevLett.120.162001}{\emph{Phys. Rev. Lett.}
  {\bfseries 120} (2018) 162001}
  [\href{https://arxiv.org/abs/1802.00349}{{\ttfamily 1802.00349}}].

\bibitem{Chen:2021azt}
X.~Chen, A.~Huss, S.~P. Jones, M.~Kerner, J.~N. Lang, J.~M. Lindert and H.~Zhang,
%  \emph{{Top-quark mass effects in H+jet and H+2 jets production}},
  \href{https://doi.org/10.1007/JHEP03(2022)096}{\emph{JHEP} {\bfseries 03}
  (2022) 096} [\href{https://arxiv.org/abs/2110.06953}{{\ttfamily
  2110.06953}}].

\bibitem{Bonciani:2022jmb}
R.~Bonciani, V.~Del~Duca, H.~Frellesvig, M.~Hidding, V.~Hirschi, F.~Moriello
  et~al., 
%  \emph{{Next-to-leading-order QCD corrections to Higgs production in
%  association with a jet}},
  \href{https://doi.org/10.1016/j.physletb.2023.137995}{\emph{Phys. Lett. B}
  {\bfseries 843} (2023) 137995}
  [\href{https://arxiv.org/abs/2206.10490}{{\ttfamily 2206.10490}}].

\bibitem{Bonetti:2020hqh}
M.~Bonetti, E.~Panzer, V.~A. Smirnov and L.~Tancredi, 
%\emph{{Two-loop mixed
%  QCD-EW corrections to $gg \to Hg$}},
  \href{https://doi.org/10.1007/JHEP11(2020)045}{\emph{JHEP} {\bfseries 11}
  (2020) 045} [\href{https://arxiv.org/abs/2007.09813}{{\ttfamily
  2007.09813}}].

\bibitem{Gao:2023bll}
J.~Gao, X.-M. Shen, G.~Wang, L.~L. Yang and B.~Zhou, 
%\emph{{Probing the Higgs
%  boson trilinear self-coupling through Higgs boson+jet production}},
  \href{https://doi.org/10.1103/PhysRevD.107.115017}{\emph{Phys. Rev. D}
  {\bfseries 107} (2023) 115017}
  [\href{https://arxiv.org/abs/2302.04160}{{\ttfamily 2302.04160}}].

\bibitem{Harlander:1998cmq}
R.~Harlander, T.~Seidensticker and M.~Steinhauser, 
%\emph{{Complete corrections
%  of Order alpha alpha-s to the decay of the Z boson into bottom quarks}},
  \href{https://doi.org/10.1016/S0370-2693(98)00220-2}{\emph{Phys. Lett. B}
  {\bfseries 426} (1998) 125}
  [\href{https://arxiv.org/abs/hep-ph/9712228}{{\ttfamily hep-ph/9712228}}].

\bibitem{Seidensticker:1999bb}
T.~Seidensticker, 
%\emph{{Automatic application of successive asymptotic
%  expansions of Feynman diagrams}},  
  in \emph{{6th International Workshop on
  New Computing Techniques in Physics Research: Software Engineering,
  Artificial Intelligence Neural Nets, Genetic Algorithms, Symbolic Algebra,
  Automatic Calculation}}, 5, 1999
  [\href{https://arxiv.org/abs/hep-ph/9905298}{{\ttfamily hep-ph/9905298}}].

\bibitem{Beneke:1997zp}
M.~Beneke and V.~A. Smirnov, 
%\emph{{Asymptotic expansion of Feynman integrals
%  near threshold}},
  \href{https://doi.org/10.1016/S0550-3213(98)00138-2}{\emph{Nucl. Phys. B}
  {\bfseries 522} (1998) 321}
  [\href{https://arxiv.org/abs/hep-ph/9711391}{{\ttfamily hep-ph/9711391}}].

\bibitem{Pak:2010pt}
A.~Pak and A.~Smirnov, 
%\emph{{Geometric approach to asymptotic expansion of
%  Feynman integrals}},
  \href{https://doi.org/10.1140/epjc/s10052-011-1626-1}{\emph{Eur. Phys. J. C}
  {\bfseries 71} (2011) 1626}
  [\href{https://arxiv.org/abs/1011.4863}{{\ttfamily 1011.4863}}].

\bibitem{Davies:2020lpf}
J.~Davies, G.~Mishima, M.~Steinhauser and D.~Wellmann, 
%\emph{{$gg\to ZZ$:
%  analytic two-loop results for the low- and high-energy regions}},
  \href{https://doi.org/10.1007/JHEP04(2020)024}{\emph{JHEP} {\bfseries 04}
  (2020) 024} [\href{https://arxiv.org/abs/2002.05558}{{\ttfamily
  2002.05558}}].

\end{thebibliography}

\end{document}